\documentclass{article}

\usepackage{arxiv}
\usepackage{amsmath,amssymb,amsfonts}
\usepackage{algorithmic}
\usepackage{graphicx}
\usepackage{subfig}
\usepackage[hyphens]{url}
\usepackage{booktabs}       
\usepackage{nicefrac}       
\usepackage{seqsplit}

\usepackage{microtype}      
\usepackage{multirow}
\usepackage{CJKutf8}
\usepackage{textcomp}
\usepackage{bmpsize}
\usepackage{xcolor}
\usepackage{lipsum}
\usepackage{enumitem}
\usepackage{xcolor}

\usepackage{listings}

\newsavebox{\mybox}
\newsavebox{\yrbox}

\usepackage[colorlinks=true,urlcolor=black]{hyperref}
\def\BibTeX{{\rm B\kern-.05em{\sc i\kern-.025em b}\kern-.08em
    T\kern-.1667em\lower.7ex\hbox{E}\kern-.125emX}}
\begin{document}

\title{SCORE: Syntactic Code Representations for Static
Script Malware Detection}


\author{
 Ecenaz Erdemir $^*$ \\
  Amazon Web Services\\
  New York, USA \\
  \texttt{ecenaz@amazon.com} \\
  \And
 Kyuhong Park $^*$ \\
  Amazon Web Services \\
  New York, USA \\
  \texttt{kyuhongp@amazon.com} \\
  \And
 Michael J. Morais \\
  Amazon Web Services\\
  New York, USA \\
  \texttt{moraismi@amazon.com} \\
  \AND
 Vianne R. Gao \\
  Amazon Web Services \\
  New York, USA \\
  \texttt{gaov@amazon.com} \\
  \And
 Marion Marschalek \\
  Amazon Web Services\\
  New York, USA \\
  \texttt{mamarsch@amazon.com} \\
  \And
 Yi Fan \\
  Amazon Web Services\\
  New York, USA \\
  \texttt{fnyi@amazon.com} \\
}

\maketitle
\def\thefootnote{*}\footnotetext{Contributed equally}\def\thefootnote{\arabic{footnote}}
\begin{abstract}
As businesses increasingly adopt cloud technologies, they also need to be aware of new security challenges, such as server-side script attacks, to ensure the integrity of their systems and data. These scripts can steal data, compromise credentials, and disrupt operations. Unlike executables with standardized formats (e.g., ELF, PE), scripts are plaintext files with diverse syntax, making them harder to detect using traditional methods. As a result, more sophisticated approaches are needed to protect cloud infrastructures from these evolving threats. In this paper, we propose novel feature extraction and deep learning (DL)-based approaches for static script malware detection, targeting server-side threats. We extract features from plain-text code using two techniques: syntactic code highlighting (SCH) and abstract syntax tree (AST) construction. SCH leverages complex regexes to parse syntactic elements of code, such as keywords, variable names, etc. ASTs generate a hierarchical representation of a program’s syntactic structure. We then propose a sequential and a graph-based model that exploits these feature representations to detect script malware. We evaluate our approach on more than 400K server-side scripts in Bash, Python and Perl. We use a balanced dataset of 90K scripts for training, validation, and testing, with the remaining from 400K reserved for further analysis. Experiments show that our method achieves a true positive rate (TPR) up to 81\% higher than leading signature-based antivirus solutions, while maintaining a low false positive rate (FPR) of 0.17\%. Moreover, our approach outperforms various neural network-based detectors, demonstrating its effectiveness in learning code maliciousness for accurate detection of script malware.

\end{abstract}


\section{Introduction}
\label{sec:introduction}
Script-based malware has emerged as a potent threat vector, frequently leveraged in attacks against Linux systems due to the versatility, portability, and ease of use of modern scripting languages. As this threat becomes more prevalent, cloud environments and Linux systems have also become prime targets, especially for script malware written in server-side languages like Python, Perl, and shell scripts. These malicious scripts can serve as standalone threats, such as denial-of-service bots, ransomware, or backdoors, or function as components in multi-stage attacks by facilitating payload delivery and execution. With capabilities comparable to traditional executable malware, including data exfiltration and system resource abuse, these script-based malware poses a significant challenge for detection mechanisms that rely on static signatures or traditional machine learning (ML) techniques.

The threat posed by script malware targeting cloud infrastructure is particularly concerning. These malicious scripts can directly access and manipulate underlying cloud resources, providing attackers with powerful capabilities. For instance, a recent Python-based credential harvester and hacking tool called Legion  and AlienFox have been observed targeting AWS console credentials, SNS, S3, and SES services \cite{Legion,AlienFox}. Attackers can leverage the hijacked cloud infrastructure for activities like mass spamming, phishing campaigns, and privilege escalation. More importantly, the prevalence of script malware attacks has surged in recent years, with reports indicating a 100\% increase since 2017 and such attacks accounting for 40\% of all cyberattacks as of 2020 \cite{Report2020}. By early 2024, the frequency had nearly doubled again over the prior two years \cite{Acronis}. Given this rapid growth and powerful capabilities of scripts \cite{snakeinfostealer,outlawbotnet,cloudcredstealer} mostly used in cloud environments, developing effective detection methods specifically targeting script malware is a critical security need.

Malware analysis broadly spans static and dynamic analysis methods. Dynamic analysis executes the malicious file in a highly-controlled ``sandbox'' environment, and observes the malicious behavior directly \cite{survey-mw-detection}. Static analysis, on the other hand, analyzes the structure, properties, and code of a malicious file {\it without executing it}, and attempts to infer that same malicious behavior from these related data. As a result, static analysis is often cheaper, safer, and faster than dynamic analysis \cite{limitsofstatic}. Typically, threat intelligence in static analysis tools---such as classical antivirus scanning products---vends as signatures, or {\it rules}, coalesced into a pattern-matching database that specifies malicious content that, if observed in a certain file, designates it `malicious'. However, such signature-based methods are often ineffective against more advanced malware that use simple code transformations to evade detection \cite{limitsofstatic,staticExe}. ML-based representation learning of malware, in contrast, presents the opportunity to learn the malicious behavior patterns to capture what static signatures cannot \cite{jain2020contrastive,UAST,Suneja2020LearningTM,feng-2020-codebert, PyMT5, Script_CNN}. However, most light-weight ML models focus on learning code behavior from unstructured byte-strings \cite{MalConv, MalConv2}, while code language models are too slow and costly due to their billions of parameters. There is a need for reasonably sized ML models that can learn more global, generalizable, invariant, robust, and/or functional patterns of malicious structure in code compared to simple byte-string patterns.

In contrast to single-format executables, analyzing scripts presents unique challenges. Scripts are often domain-specific, requiring specialized knowledge to extract features and model its behavior accurately for every specific script language. As the feature extraction becomes more sophisticated, complexity increases. On the other hand, byte-strings alone might not contain as much structured information as well-designed features. In this paper, we propose feature extraction methods that directly target these structures of scripts and detection methods that aim to understand the context of these scripts.

In this paper, we address these research questions:
\begin{itemize}
    \item \textbf{RQ1}: How can code from server-side programming languages be effectively represented as features for scalable malware detection using deep learning (DL)? Which feature representation method provides the most useful information to detect malicious behavior?

    \item \textbf{RQ2}: Can DL-based models learn the complex structure of scripts, and which types of DL models most accurately identify malicious code behavior?

    \item \textbf{RQ3}: Does the best-performing DL-based model surpass conventional rule-based malware detectors in accuracy and threat coverage when applied to scripts?
\end{itemize}

\begin{figure*}[pt]
\center
    \includegraphics[width=0.9\textwidth]{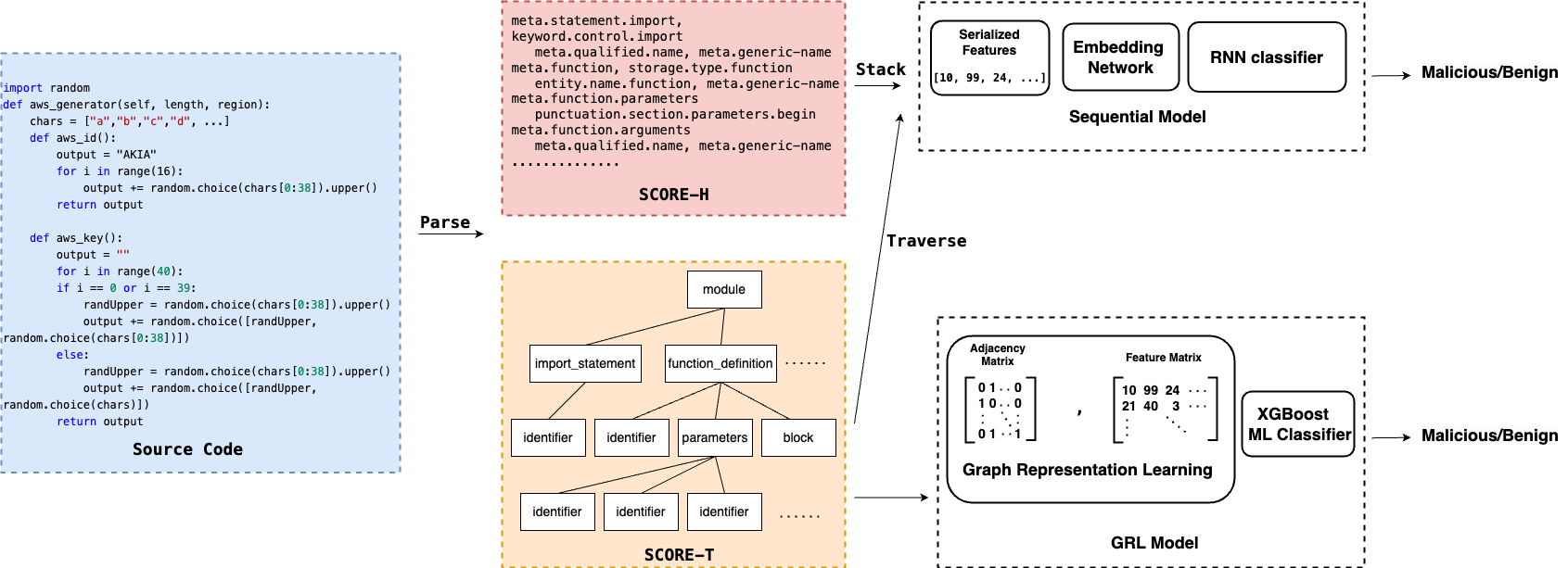}
    \caption{System overview: Our system first parses scripts for their SCORE-H or SCORE-T features. In the case of SCORE-H, our SM is fed with the serialized input and generates malicious or benign verdict. In the case of SCORE-T, ASTs are either traversed and fed to the SM or fed directly to a GRL model along with their adjacency matrix to output a malicious or benign verdict.}
    \label{fig:system_overview}
\end{figure*}

To address these questions, firstly, we propose script malware detection methods for server-side languages by leveraging popular code parsing libraries to extract features from the plain-code. We present two approaches for representing code as features: SCORE-H which is based on syntactic code highlighting and SCORE-T which is based on abstract syntax trees (ASTs). While SCORE-H parses the keywords in a sequentially hierarchical level, SCORE-T parses a program's syntactic structure into a hierarchical tree representation. Both approaches pair their syntactic structure features with raw byte-strings of the scripts for more information-rich representations. Secondly, we propose malware detection models: a sequential model (SM) and a graph representation learning (GRL)-based model. SM contains multiple layers of convolutional neural networks (CNNs) to extract hierarchical embeddings from SCORE-H features, followed by a recurrent neural network (RNN), e.g., bi-directional long short-term memory networks (bi-LSTM). Moreover, we propose a variation of SM with simpler CNN embeddings for serialized SCORE-T features. Finally, our GRL-based model leverages graph embeddings obtained by graph similarity learning and contains a ML-based detector, such as XGBoost \cite{XGBoost}, for detecting static malicious behavior. System overview of our proposed approaches is shown in Figure \ref{fig:system_overview}, where each end-to-end model represents a malware detector, e.g., SM with SCORE-H features, SM with SCORE-T features and GRL model with SCORE-T features.

We compare the proposed approaches against commercial antiviruses (AVs), a byte-level feature-based malware detector \cite{MalConv2}, a sequential and a graph-based neural network malware detector that utilize ASTs \cite{UAST}, and finally an ML detector that utilizes embeddings from CodeBERT, a multi-lingual foundation model pre-trained on Natural Language (NL) - Programming Language (PL) pairs \cite{feng-2020-codebert}. Our SM shows significant performance for serialized features due to the sequentially written form of scripts. Serialization of the tree structure is highly significant and determined by the traversal method, such as breadth first traversal (BFT) and depth first traversal (DFT). When we have tree/graph structured features and access to threat labels during training, GRL-based model improves the detection performance of the SM for DFT. On the other hand, SM with BFT serialization shows the best performance of all the methods considered in this paper. Overall, all of our proposed approaches address the limitations of AVs, byte-level and AST-based approaches as well as token-based approaches in the literature by leveraging advanced code parsing capabilities and deep learning (DL) techniques to detect script malware attacks.

\sloppy
Our contributions can be summarized as follows:
\begin{enumerate}
\item We introduce novel feature extraction techniques tailored for server-side script languages, including a syntactic highlighting-based extractor that represents code functionality as a sequence and an AST-based extractor that captures deeper understanding of code through hierarchical syntactic representation. These extractors serve as programming language tokenizers and are integrated with a sequential neural network, enabling an understanding of code rather than relying solely on pattern matching techniques. We further incorporate this hierarchical structure into embeddings by GRL. To the best of our knowledge, these approaches are novel in the realm of malware detection.
\item We have curated a comprehensive collection of malicious and benign scripts. Our extensive evaluations, coupled with comparisons to existing methodologies, demonstrate that our approaches:
\begin{itemize}
    \item Outperform a commercial AV, an open source AV, ML-based byte-level malware detectors, AST-based sequential and graph neural network detectors, and a pre-trained CodeBERT-based malware detector.
    \item Provide coverage for more than 95\% of high-priority threats, including cryptominers, ransomware, and credential stealers.
\end{itemize}
\end{enumerate}

\section{Motivation and Scope}

\subsection{Motivation}
To illustrate and motivate our problem clearly, we consider a real-world script malware case study. Recent Python-based credential harvester and hacktool, called Legion, exploits web servers that run Content Management Systems (CMS), PHP, or PHP-based frameworks like Laravel. It employs regex patterns to extract credentials from these servers for various web services such as email providers, cloud services like AWS, server management systems, databases, and payment systems like Stripe and PayPal \cite{Legion}. More specifically, Legion targets AWS console credentials, AWS SNS, S3 and SES specific credentials. It uses the infrastructure of the hijacked service for mass spamming or opportunistic phishing campaigns, and also includes code to implant webshells, brute-force CPanel or AWS accounts and send SMS to dynamically-generated US mobile numbers. Similarly, another real-world credential harvester, called AlienFox, is used by attackers to extract sensitive information such as API keys and secrets from service providers such as AWS \cite{AlienFox}. AlienFox can establish AWS account persistence and privilege escalation, collect send quotas and automate spam campaigns through victim accounts or services. Hence, it is crucial to accurately detect these malicious scripts for cloud users' security. 

Traditional signature-based scanners rely on identifying specific patterns to detect such threats. For instance, in the source code in Figure \ref{fig:system_overview}, we showcase a function from Legion that uses brute force to generate AWS ID and KEY by randomly selecting characters to create strings. Attackers can easily evade detection by modifying function names (e.g., aws\_id and aws\_key ), using third-party libraries instead of built-in ones, or obfuscating values and strings. While these evasion techniques often bypass pattern matching approaches in traditional AV software, syntax highlighting and abstract syntax tree-based feature representations of the code remain unaltered. This is because these features capture the syntactic structure of the code, which remains consistent even when the byte-strings are modified. Therefore, our motivation is to capture the meaningful syntactic structure of the code by our proposes feature extraction methods and learn the static behavior of the malicious code through advanced DL techniques for successful script malware detection.

\subsection{Scope}
In this work, we concentrate exclusively on server-side script languages within the context of cloud security, such as Python, Perl and Bash. By "server-side scripts", we refer to programs operating on servers, whether they're web servers, application servers, etc. These scripts handle requests, execute computations, interact with databases or file systems, and generate responses or outcomes for clients or other server-side components. They serve purposes such as automating server provisioning and configuration, managing deployments, orchestrating services, processing and analyzing data, implementing serverless functions or microservices, and facilitating communication between different server-side components. These are the most common tasks in cloud environment, therefore targeted by most threat actors. Orthogonal to existing works that consider the client-side languages, e.g., Javascript, PHP, HTML, etc., our primary focus is on identifying and mitigating malware threats for server-side covering Bash, Python and Perl languages. 

It is important to note that our script malware detection method, which relies on code structure, does not address Supply Chain Attacks, such as the 2021 Python Package Index (PyPI) attack \cite{PyPI}, where malicious code was distributed through the official PyPI by compromising the accounts of package maintainers. These attacks are not directly related to script code itself and are out of scope of this paper \cite{SupplyChain1,SupplyChain2}.

\subsection{Related Work}
In the literature, there are various studies that encompass client-side script languages, most specifically tailored for JavaScript \cite{JS_detection1, JS_detection2, JS_detection3, JS_detection4, JS_detection5}. On the other hand, server-side script malware detection has not been well-studied in the context of DL. In this paper, we propose script malware detection for server-side languages utilizing syntactic feature extraction and DL-based detection techniques. Our approaches stand as a novel mitigation for such threats targeting cloud environment. This section briefly summarizes the related work to our paper.

In recent years, there has been significant interest in using DL for malware detection by converting malware binaries into grayscale images \cite{prajapati2021empirical,Pant2022Image,Freitas2022Malnet,sang2018malware,Agarap2017TowardsBA}. However, the focus has primarily been on malware binaries since it is easier to find visual patterns in binaries due to their fixed and regular structure unlike scripts. Existing approaches typically convert these binaries into 8-bit vectors, treating them as grayscale images. This has enabled the application of computer vision techniques. However, scripts lack such fixed, simple structures by nature, and hence, these approaches are out of scope of this paper. Instead, script malware requires a contextual understanding of code rather than focusing on visual appearance in byte-strings, which is the approach we propose in this work.

MalConv \cite{MalConv} is a well-known DL-based malware detector that uses raw byte sequences as input without pre-processing. Its CNN-based detection model can process a raw byte sequence of over two million steps, and allow for interpretable sub-regions of the binary to be identified. However, its linear complexity dependence on the sequence length is memory-inefficient. Later, MalConv2 \cite{MalConv2} improved \cite{MalConv} by introducing a temporal max pooling approach which is much more memory efficient and faster to train model. This model has two modules, one for capturing overall context and one for extracting detailed code features, i.e., content. It consists of $1D$ CNNs and gating for feature weighting. The extracted context and features are fed to an attention mechanism that weights the features according to their context. This aspect of MalConv2 makes it a general purpose malware detector. Despite being cheap and generalizable, byte-level detection has limitations in capturing sophisticated malicious behavior, and often obfuscation or commented byte-strings can cause malware evasion.

As an effort to extract structural information from scripts, as opposed to byte analysis, Dendroid approach \cite{Dendroid} introduces a notion of code chunks as a functional unit of static malware analysis, which inspired our novel feature extraction method explained in Section \ref{sec:SSH_based_features}. Dendroid uses a hierarchical—dendrographic—clustering to group code chunks by function, and even common source code. It replaces a sequence of instructions in a code chunk by a list of statements defining its control flow. After parsing each code chunk with a particular grammar for Android malware, the resulting structure is a sequence of symbols of varying length. These hierarchical sequences of symbols help representing code as a natural language and employ language models for downstream tasks.

Another method for code representation learning is proposed in \cite{UAST}, where ASTs of programming languages are considered. First a unified vocabulary is created for different AST nodes with the same functionality, then two AST representation learning approaches are proposed and combined together for improved performance. These approaches are namely 1) Sequence-based AST network (SAST), 2) Graph-based AST network (GAST) and 3) Unified AST network (UAST) which is their joint model. Structured AST representations of scripts are fed to a graph convolutional network (GCN) in GAST, and in parallel, their serialized version is fed to a bi-LSTM network in SAST. Outputs of both networks are joined in UAST approach for task classification, such as code smell classification, defects classification, etc. However, both bi-LSTM network and GCN of \cite{UAST} use relatively simple networks, e.g.,single hidden layers, flattened tree structure for graph convolutions and etc. These networks are nor sufficient in learning complex malicious behavior.

A state-of-the-art (SotA) graph similarity learning approach is proposed in \cite{GraphMatching}, which studies similarity learning problem specifically for control-flow-graphs (CFG). Motivated by binary function similarity search, a given binary is checked whether it contains a code with a known vulnerability from a database by checking the control-flow-graph similarity. In \cite{GraphMatching}, GNNs are utilized to embed graphs into a vector space, and represent similar graphs close and dissimilar graphs apart in the vector space.

CodeBERT \cite{feng-2020-codebert} is a popular pre-trained model that captures the syntax and meaning of both natural language (NL) and programming languages (PL). By leveraging the BERT architecture, CodeBERT performs tasks like code search, summarization, and completion. Trained on a large GitHub corpus using masked language modeling, it achieves state-of-the-art results in various code intelligence benchmarks. While CodeBERT excels in understanding the textual representation of code, task oriented and more structured approaches, such as AST-based methods, can provide a deeper understanding by directly modeling the syntactic structure of code, which can be crucial for tasks requiring precise functional interpretation.

Our approach for code representations and learning static malicious behavior both sequentially and in graphical structure are inspired by the related work. However, these do not consider our application use-case. They either propose task detection for benign code, which considers simpler programs than malicious code specialized to evade detection, or more elaborate CFG-based reverse engineering. While transformer-based representations increase latency, they don't offer as significant an improvement as incorporating syntactic structure into the features, which is justified in our results. In this paper, we directly target detecting static malicious behavior in scripts, and show that our approaches significantly outperform aforementioned work in the context of script malware detection.

\begin{figure*}[pt]
  \centering
  \includegraphics[width=0.9\textwidth]{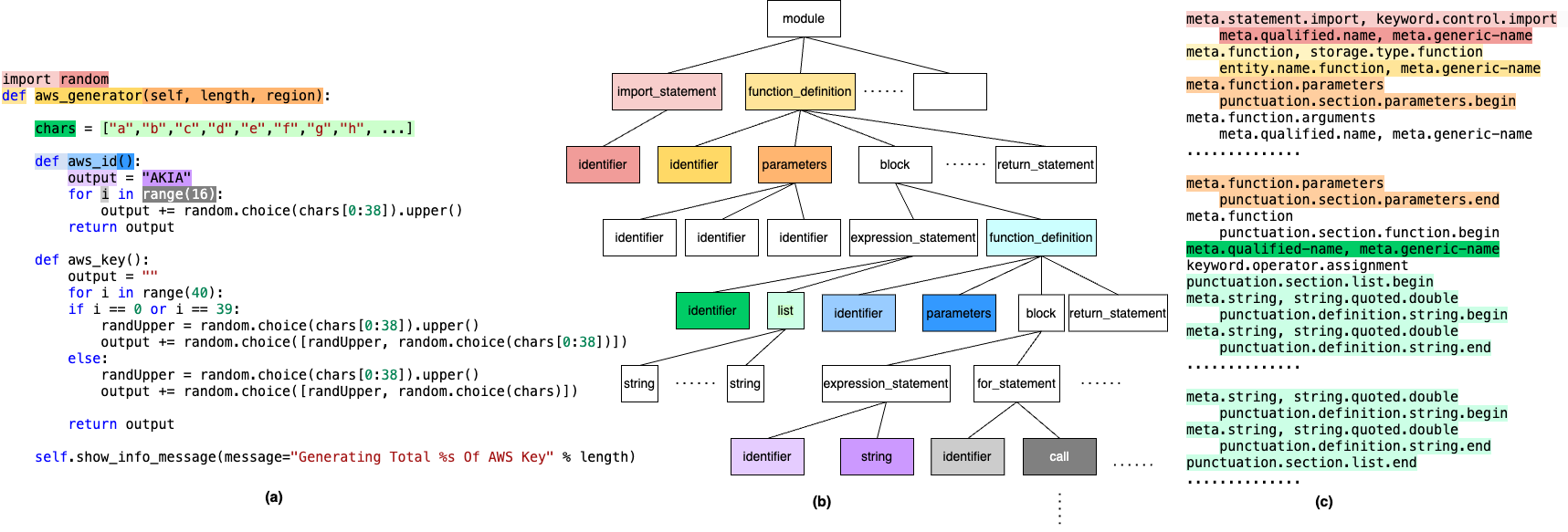}
  \caption{Exemplars of (a) Python script from Legion, a credential harvester, (b) its abstract syntax tree and (b) syntactic code highlighting expressions.}\label{fig:ssh_ast_example}
\end{figure*}
\section{Syntactic COde REpresentations (SCORE)}
\label{sec:semantic_code_representations}
Specialization to {\it script} malware detection allows us to benefit from existing highly sophisticated tools for parsing plain-text code. Static code analysis is a fundamental task for programming language design, and, as such, we can directly utilize tools like syntax highlighting, used for visualization by humans, and abstract syntax tree construction, used for compilation or interpretation by computers. Both of the above use complex {\it grammars} that efficiently construct various representations of the input code. Our contributions translate intermediate data structures of such parsing libraries into features that can be harnessed as input for ML-based malware detection. We describe the methodology of our syntax {\bf H}ighlighting-based feature extractor (SCORE-H) and our abstract syntax {\bf T}ree-based feature extractor (SCORE-T) in the following subsections (see Figure \ref{fig:ssh_ast_example}).

In this paper, we focus our static analysis on server-side languages Python, Perl and Bash / Shell scripts, which are directly executable by attackers and target the server-side, and for which we observe high volumes of malware samples. Other source code languages ({\it e.g.} C, C++, Java) must be compiled first, and are generally not used in malware attacks---the resulting executables are used instead. As a result, those executables are out-of-scope for script malware detection, and the source code will not be observed as ``malware'' on victim machines.

\subsection{Syntax highlighting features (SCORE-H)}
\label{sec:SSH_based_features}
\sloppy
Syntax highlighting is a mainstay of modern IDEs that colorizes keywords, variable names, class methods, function definitions, brackets, indents and more to enhance user experience. These colorizations encode the functional role and context of code strings, as well as operate at near-zero latency to highlight code as it is written by the user. To harness these ideas for malware detection, we adapted a popular syntax highlighter syntect \cite{syntect}, which is used in text editors such as VSCode \cite{VSCode} and Sublime Text \cite{Sublime}. We chose syntect primarily because it is a fast and reliable Rust-based syntax highlighter. Additionally, it is already utilized in production in industry and numerous open-source projects \cite{syntect}.

\sloppy
Specifically, we use only the parsing capabilities of the library, which use the recursive regex grammars also used in Sublime Text to tokenize plain-text code into named regex matches called {\it scopes} (Figure \ref{fig:ssh_ast_example}b). Scopes are hierarchically organized identifiers of syntactic content, composed of dot separated {\it atoms}. For example, in Python code \texttt{`import base64'}, \texttt{import} keyword matches the scopes \texttt{meta.statement.import.python} and \texttt{keyword.control.import.python}, signposting its role as the keyword of an import statement. In practice, we truncate the last atom identifying the language (\texttt{python}) to condense the syntactic features and permit cross-language inferences in the model. However, traces of each script's language remain, as keywords, syntax, and usage still vary across languages.

We parse each script into a {\it stack} of such scopes, and unravel that stack into a sequence of tokenized byte-strings and their corresponding scopes. Such coupled sequences of scope and raw-byte content is the representation defining the SCORE-H feature extractor. We only utilize this feature representation in sequential models (see Section \ref{sec:hierarchical_cnnrnn_model}).

\subsection{Abstract syntax tree features (SCORE-T)}
\label{sec:AST_based_features}
ASTs are a core code representation for compilers to generate intermediate code, linters to detect errors, and more. The tree structure of nested nodes and edges reflects hierarchical syntactic structure and context of the corresponding code, and parses with very low latency. To harness these ideas for malware detection, we adapted the AST parsing library tree-sitter \cite{tree-sitter}. Our primary reason for choosing Tree-sitter is its Rust-based implementation, similar to Syntect, ensuring it is fast, robust, and secure. Additionally, Tree-sitter is dependency-free and can be seamlessly embedded in any application. 

Tree-sitter has its own context-free regex grammars for parsing various code languages into ASTs, and these tokenize plain-text code into named nodes (Figure \ref{fig:ssh_ast_example}a). For example, in the same Python code above \texttt{`import base64'}, the \texttt{import} keyword generates the node \texttt{import statement}, signposting its role as the keyword of an import statement. The subsequent \texttt{identifier} node for the imported package, \texttt{base64}, becomes a leaf branching from that node. In this way, ASTs nest functionally-related chunks of code, such as function definitions and control-flow statements, into subtrees. (Syntax highlighting, in contrast, only partially replicates this behavior.) In practice, we further standardize the node naming schema into a unified cross-language vocabulary, following the unified AST approach proposed in \cite{UAST}. For example, the root node of Bash is represented by $program$, while in Python and Perl ASTs they are represented by $module$ and $source$ $file$ nodes, respectively. Despite their different names, they represent the same concept: the code file or program. Additionally, the node names for code blocks also vary between languages, with $compound$ $statement$ in Perl and $block$ in Python. We unify the node naming to condense the AST features and permit cross-language inferences as above, noting the same limitations.

We parse each script into an AST and couple each node with its associated raw code (byte-strings), and use this graph data structure as features by either ({\it i}) traversing the graph and generating a sequence or ({\it ii}) directly utilizing the graph. In the former case, we traverse the AST to generate a coupled sequence of AST node-names and raw-byte content, using either depth first traversal (DFT) and breadth first traversal (BFT), and can apply sequential models (Section \ref{sec:hierarchical_cnnrnn_model}). In the latter case, we preserve the tree structure, and apply graph-based models (Section \ref{sec:graph_based_ast_model}).

\section{Malware detection}
\label{sec:malware_detection}
In this section, we propose sequential and graph representation learning-based models (SM and GRL, {\it respectively}) that translate our syntax highlighting and AST-based code representations into malicious and benign verdicts. We designed the SM particularly for SCORE-H features due to their sequential hierarchical structure, and discuss how we adapt it for serialized SCORE-T features. We designed the GRL approach particularly for SCORE-T features, which uses graph similarity learning on the intrinsic graph structure of those features. Both SM and GRL are designed specifically for their corresponding features to learn the static malicious/benign behavior of scripts.

\begin{figure*}[pt]
  \centering
  \includegraphics[width=0.9\textwidth]{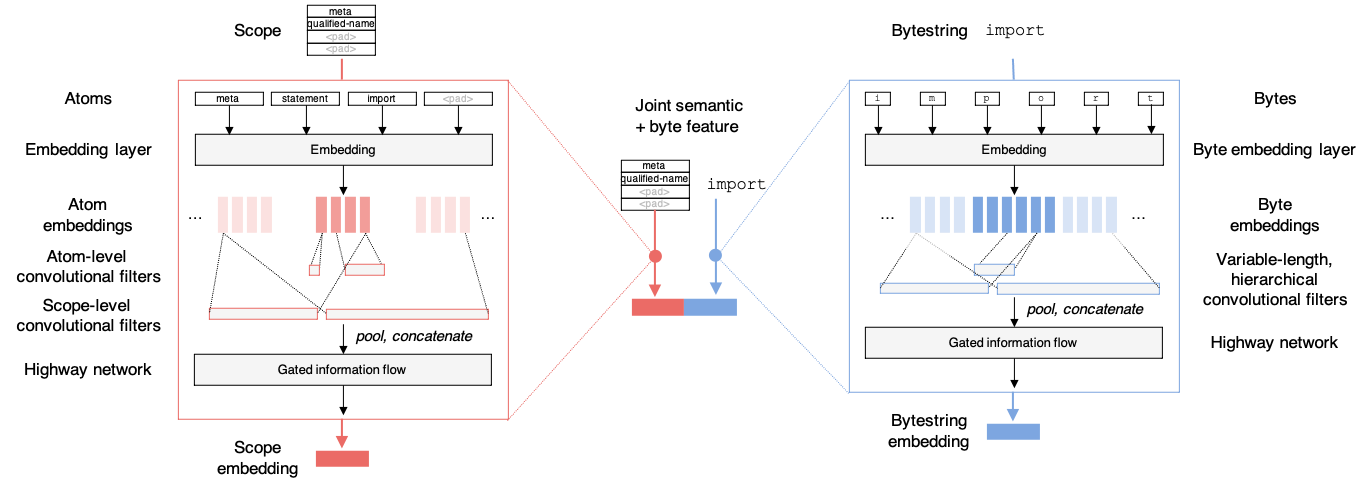}
  \caption{Syntax highlighting ({\it left}) and byte ({\it right}) feature embedding module for SCORE-H model. The SCORE-T model uses the same byte-string embedding submodule, but only a subset of the convolutional filters shown in the scope (node-name, {\it resp.}) embedding submodule. }\label{fig:CNN-RNN_features}
\end{figure*}

\subsection{Sequential model (SM)}
\label{sec:hierarchical_cnnrnn_model}
The sequential script malware detector is a convolutional and RNN model of the joint byte-syntactic features of both the SCORE-H and serialized SCORE-T feature representations. It is composed of two modules---one for generating embeddings of the joint byte-syntactic features, and one for generating malicious and benign verdicts from the resulting sequences of feature embeddings.

\sloppy
The embedding module fits two embeddings, one for the syntactic features (syntax highlighting and AST-based features) and one for the byte features, and concatenates them together into a joint embedding per item in the sequence (Figure \ref{fig:CNN-RNN_features}). It is common to implicitly draw analogies to language modeling when designing embeddings of tokenized sequential data, {\it e.g.} to sequences of letters defining words, and sequences of words as documents; to this end, we draw particular attention the unusual hierarchical structure of the scopes and atoms in the SCORE-H features. We recognized that such structure parallels not the organization of Western languages where Letter$<$Word$<$Phrase$<$Sentence, but instead the organization of East Asian languages using hanzi/kanji where Radical$<$Character$<$Word$<$Phrase$<$Sentence. The decomposition of {\it scopes} into {\it atoms}, {\it e.g.} \texttt{meta.statement.import}$\rightarrow$[\texttt{meta}, \texttt{statement}, \texttt{import}] , is analogous to the decomposition of {\it characters} into {\it radicals}, {\it e.g.} \begin{CJK*}{UTF8}{goth}語$\rightarrow$[言,五,口]\end{CJK*}, insofar as each radical/atom encodes distinct-but-related information to the aggregate character/scope. This structure hard-codes similarity between scopes with shared atoms irrespective of model training, such that \texttt{meta.statement.import} is similar to \texttt{meta.statement.try} and also to \texttt{keyword.control.import}.

\begin{figure*}[ht]
  \centering
  \includegraphics[width=0.9\textwidth]{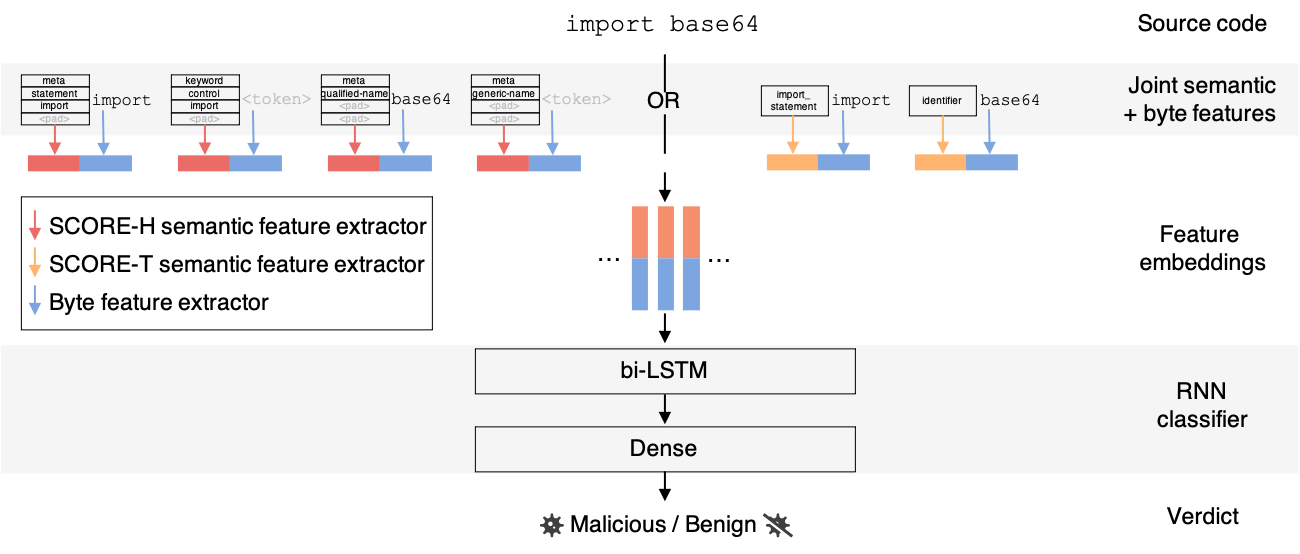}
  \caption{SM architecture. Joint byte-syntactic features are extracted from syntax highlighting intermediates for SCORE-H (red, {\it left}) or ASTs for SCORE-T (orange, {\it right}). For either featureset, embeddings of syntactic features and bytes are concatenated into a sequence of inputs to a bi-LSTM RNN classifier, which yields malicious or benign verdicts.}\label{fig:CNN-RNN}
\end{figure*}

Operationalizing this analogy, we designed a specialized CNN-based model adapted from \cite{CNN-RNN} to generate embeddings of the SCORE-H syntax highlighting features (see Figure \ref{fig:CNN-RNN_features}, {\it left}). To capture both atom- and scope-level meaning in the trained embeddings, we use three 1-D convolutional modules with filter dimensions $64, 128, 192$ and kernel sizes from $1$ to $3$ (covering one atom/radical) and three modules with kernel sizes from $6$ up to $18$ (covering three scopes/characters). Then, the concatenation of all convolutional module outputs is sent through a highway layer \cite{highway}, which implements a self gating mechanism to control information flow and adds skip connections to prevent vanishing gradients. A plain feedforward neural network layer typically applies an affine transform followed by a nonlinear activation, $H(\cdot, \mathbf{W_H})$, to its input $\mathbf{x}$, e.g.,
\begin{equation}
    \mathbf{y}=H(\mathbf{x},\mathbf{W_H}),
\end{equation}
where $\mathbf{y}$ is the layer output and the transform $H$ is parametrized by $\mathbf{W_H}$. On the other hand, a highway layer contains two additional nonlinear transforms $T(\mathbf{x},\mathbf{W_T})$ and $C(\mathbf{x},\mathbf{W_C})$ such that:
\begin{equation}
    \mathbf{y} = H(\mathbf{x}, \mathbf{W_H}) \cdot T(\mathbf{x},\mathbf{W_T}) + \mathbf{x} \cdot C(\mathbf{x},\mathbf{W_C}).
\end{equation}
We designed a similar CNN-based model to generate embeddings of the SCORE-H byte features (see Figure \ref{fig:CNN-RNN_features}, {\it right}). The only difference is we learn an embedding for each byte, followed by three 1-D convolutional modules with filter dimensions $64, 128, 192$ and kernel sizes $16, 32, 128$, respectively. These two embeddings are concatenated into a joint byte-syntactic embedding. 

We designed a simplified adaption of this joint embedding module for SCORE-T features, which can serialized into sequences by breadth first traversal (BFT) or depth first traversal (DFT). We highlight that hierarchical atom- and scope-level convolutional filters, as designed for SCORE-H, do not apply to the non-hierarchical nodes of SCORE-T. Therefore, we modify the embeddings by applying only the ``scope''-level convolutional filters ({\it cf.} Figure \ref{fig:CNN-RNN_features}) to the tree nodes of SCORE-T features. We use the same byte embeddings as-is. We model the resulting sequence of joint byte-syntactic SCORE embeddings using a bi-LSTM with $2$ layers and hidden dimension of $256$, followed by an attention module and an output dense layer to classify the script as malicious or benign (Figure \ref{fig:CNN-RNN}). The model is trained by optimizing the binary cross entropy (BCE) loss between the true labels and predictions:
\begin{equation}
    L_{BCE}=-\frac{1}{N} \sum^{N}_{i=1} \big( y_i \cdot \log(\hat{y}_i) + (1-y_i) \cdot \log(1-\hat{y}_i) \big),
\end{equation}
where $y_i$ is the true label of $i^{th}$ sample and $\hat{y}_i$ is the prediction of SM for $i^{th}$ sample. In our model, the bi-LSTM acts as a sequential classifier that detects global features that are informative of maliciousness. Overall, our approach utilizes the unique structural characteristics of code representations to create a robust script malware detector. The combination of CNN and RNN allows for encoding of both local and global features, enabling our model to capture the complex relationships between atoms, scopes, and code chunks.

\subsection{Graph representation learning (GRL) model}
\label{sec:graph_based_ast_model}
This section presents a methodology that use byte-syntactic representations of scripts by incorporating their tree structure into the detection model. We leverage the graph similarity learning approach proposed in \cite{GraphMatching} to obtain graph embeddings from our tree-structured SCORE-T features. Our goal is to learn embeddings for scripts such that similar scripts have similar graph vector representations, and these models are invariant to permutations of graph elements. Our GRL-based model computes graph node representations through a propagation process that iteratively aggregates local structural information, which is used directly for the classification of malicious behavior. We treat SCORE-T features as graphs, and apply GRL to learn embeddings. The embedding models learn vector representations of ASTs that makes similar graphs close in the vector space, such as benign-benign and malicious-malicious, and dissimilar graphs far apart, such as benign-malicious scripts. 

Given two graphs $G_1=(V_1,E_1)$ and $G_2=(V_2,E_2)$ represented by their sets of nodes $V_*$ and edges $E_*$, we aim to compute a similarity score $s(G_1,G_2)$ between these two graphs. In this context, we consider our ASTs as the graphs, and each node $i \in V$ in the graph is associated with a vector $x_i$, and each edge $(i,j) \in E$ is  associated with a feature vector $x_{i,j}$. If a node or edge does not have any associated features, we set the corresponding vector to a constant vector of $1$s. Since our features involve both syntactic features (syntax highlighting and AST-based features) and their corresponding byte features, we represent $x_i$’s as a concatenation of byte-syntactic features. Since we do not have edge features in our SCORE-T features, we set them to a constant vector of $1$s.

The embedding model represents each graph as a vector, and uses a similarity metric in that vector space to measure the similarity between graphs. This model comprises of three parts: encoder, propagation layers and an aggregator \cite{GraphMatching}. The encoder maps the node and edge features to initial node and edge vectors through separate multilayer perceptrons (MLPs):
\begin{align}
    \mathbf{h}_i^{(0)} &= MLP_{node}(x_i), \forall i \in V \\
    \mathbf{e}_{i,j} &= MLP_{edge}(x_{i,j}), \forall (i,j) \in E.
\end{align}
Then, a propagation layer maps a set of node representations $\{\mathbf{h}_i^{(t)}\}_{i \in V}$ to new node representations $\{\mathbf{h}_i^{(t+1)}\}_{i \in V}$, as follows:
\begin{align}
    \mathbf{m}_{j \rightarrow i} &= f_{message}(\mathbf{h}_i^{(t)},\mathbf{h}_j^{(t)},\mathbf{e}_{i,j}) \\
    \mathbf{h}_i^{(t+1)} &= f_{node}\Big(\mathbf{h}_i^{(t)},\sum_{j:(j,i)\in E}\mathbf{m}_{j \rightarrow i}\Big).
\end{align}
Here, $f_{message}$ is an MLP on the concatenated inputs, and $f_{node}$ is a gated recurrent unit (GRU). To aggregate the messages, we can use sum, mean, max or an attention-based weighted sum. Throughout multiple layers of propagation, the representation for each node will accumulate information in its local neighborhood. After $T$ rounds of propagation, an aggregator takes the set of node representations $\{\mathbf{h}_i^{(T)}\}$ as input, and computes a graph level representation $\mathbf{h}_{G}=f_G\big(\{\mathbf{h}_i^{(T)}\}\big)$. We use the following aggregation module,
\begin{align}
   \hspace{-0.15cm} \mathbf{h}_G \hspace{-0.1cm} = \hspace{-0.1cm} MLP_G \hspace{-0.05cm} \Bigg( \hspace{-0.05cm} \sum\limits_{i \in V}\sigma \hspace{-0.05cm} \Big( \hspace{-0.1cm} MLP_{gate}(\mathbf{h}_i^{(T)}) \hspace{-0.1cm} \Big) \hspace{-0.1cm} \odot \hspace{-0.1cm} MLP(\mathbf{h}_i^{(T)}) \hspace{-0.1cm} \Bigg),
\end{align}
which transforms node representations and then uses a weighted sum with gating vectors to aggregate across nodes. The weighted sum can help filtering out irrelevant information and works significantly better empirically. After the graph representations $\mathbf{h}_{G_1}$ and $\mathbf{h}_{G_2}$ are computed for the pair $(G_1,G_2)$, we compute their similarity using the Hamming similarity metric. This metric outperformed Euclidean and cosine similarity for our dataset.

Contrastive representation learning creates an embedding space where similar sample pairs stay close while dissimilar ones are far apart. In our approach, we consider label-wise (malicious-benign) and threat-wise (threat types) similarities. We describe our loss below, which is then optimized with gradient descent. For applications where it is necessary to search through a large database of graphs with low latency, it is beneficial to have the graph representation vectors be binary, i.e. $\mathbf{h}_G \in \{-1,1\}^H$, so that efficient nearest neighbor search algorithms may be applied. In such cases, we can minimize the Hamming distance of positive pairs and maximize it for negative pairs. To achieve this we pass the $\mathbf{h}_G$ vectors through a $tanh$ transformation, and optimize the following pair loss:
\begin{align}
    L_{pair}=\mathbb{E}_{(G_1,G_2,\ell)}[(\ell-s(G_1,G_2))^2]/4,
\end{align}
where $s(G_1,G_2)=\frac{1}{H}\sum^H_{i=1}tanh(h_{G_1i}) \cdot tanh(h_{G_2i})$ is the approximate average Hamming similarity. The loss is bounded in $[0,1]$, and it pushes the positive pairs to have Hamming similarity close to $1$, and negative pairs to have similarity close to $-1$. This loss is found to be more stable than Euclidean distance.
After learning the embeddings, we use an XGBoost \cite{XGBoost} classifier for malware detection from these embeddings.

\section{Experimental Study}
In this section, we present our experiments and evaluate their success with respect to the following goals:
\begin{enumerate}
    \item \textbf{High Accuracy} Our first goal is to demonstrate the high accuracy of our proposed methods, measured in terms of True Positive Rate (TPR), False Positive Rate (FPR), Precision, F1 score, and AUROC score.
    \begin{itemize}
        \item We established the accuracy of our experiments based on curated ground-truth data from reliable sources including VirusTotal (VT) and an in-house honeypot (HP) for malicious files.
        Benign scripts were collected from newly provisioned cloud instances and public GitHub repositories, simulating a diverse and realistic cloud user environment.
        Our dataset incorporates a mix of large and small files, including obfuscated ones, mirroring common scenarios faced by cloud users (Section \ref{subsec:Dataset}).
        \item Our approach aims for a low FPR and high TPR accuracy regime, aligning with the low tolerance of the malware domain to False Positives (FPs) and False Negatives (FNs). We achieved a detection rate of up to $0.98$ with a minimal FPR of $0.00172$ (Section \ref{subsec:Overall}).
        \item We conducted manual FP/FN analyses to understand the reasons for missed detections. This analysis focuses on identifying patterns common in real-world user environments and assessing the likelihood of users encountering similar threats. (Section \ref{subsec:FPFN})
    \end{itemize}
    \item \textbf{Comparison with Existing Tools} Another goal is to evaluate the superiority of our approaches compared to existing tools.
    \begin{itemize}
        \item BitDefender is a commercial AV, while ClamAV is an open-source AV toolkit both utilizing signature-based malware detection. They target a wide range of file-types and malware families. We aim to show that these AVs are insufficient for script malware detection and showcase significant performance improvements with our methods. (Section \ref{subsec:ComparisonAV})
        \item Our approach includes various feature extraction methods and deep learning-based approaches, therefore we compare our approaches both in terms of preprocessing and DL-based detection models. We compare our approaches with MalConv2 which originally operates at the byte-level and leverages CNNs for content and context extraction from files. We also do an ablation study on the proposed AST-based features and train MalConv2. The objective is to showcase the superiority of our feature extraction over byte-level representations as well as our model architecture over the CNN approach of MalConv2 in a fair comparison. Furthermore, we conduct comparisons with AST-based code representation learning approaches. This comparative analysis aims to highlight that despite utilizing similar feature extraction techniques, our DL-based detectors exhibit a more profound understanding of scripts compared to existing code representation learning methods. Finally, we compare our approaches with a token-based (CodeBERT) pre-trained code representation learning approach. (Section \ref{subsec:ComparisonDL})
    \end{itemize}
    \item \textbf{High Threat Coverage} Finally, we aim to achieve comprehensive coverage of threats commonly encountered in cloud environments that pose risks to users. As threat identification ground-truth, we utilized both BitDefender and VT for a fair evaluation of our best approach. (Section \ref{subsec:ThreatCoverage})
\end{enumerate}
By addressing these goals, we demonstrate the effectiveness and superiority of our script malware detection methods in realistic cloud environments.

\begin{table}[pt]
\centering
\caption{Dataset summary.}
\label{tab:dataset}
\setlength{\tabcolsep}{8pt}
\begin{tabular}{l|c|c|c|c|c|c|c|c|c}
\toprule
\textbf{Lang.}   & \multicolumn{3}{c|}{\textbf{Benign Set}} & \multicolumn{3}{c|}{\textbf{Malicious Set}} & \multicolumn{3}{c}{\textbf{Balanced Set}} \\ \cmidrule{2-10}
   & EC2        & GitHub     & Tot.  & VT    & HP. & Tot. & Train & Val. & Test \\
        \midrule
        Python         & $56432$    & $278658$   & $335090$  & $15991$ & $99$ & $16090$  & $26816$ & $2682$ & $2682$ \\ \midrule
        Bash               & $7604$     & $20054$  & $27658$  & $21767$  & $9152$  & $30919$  & $46100$ & $4608$ & $4608$ \\
        \midrule
        Perl               & $1005$      & $245$   & $1250$ & $1125$  & $125$ & $1250$  & $2084$ & $208$ & $208$  \\
        \bottomrule
\end{tabular}
\end{table}

\subsection{Dataset}
\label{subsec:Dataset}
We curated a balanced dataset from our total script malware and benign-ware in Python, Bash, and Perl, as detailed in Table \ref{tab:dataset}.
The dataset was split into train/validation/test sets with a 10:1:1 ratio (75K:7.5K:7.5K samples), maintaining equal representation of benign and malicious files across languages. Additional benign files were preserved for further analysis in Appendix \ref{apx:language_coverage}. 

\textbf{Benign Set}: Scripts were collected from newly provisioned cloud instances such as EC2 (including Amazon Linux 2 and Ubuntu), and GitHub repositories with at least 1000 stars. This approach ensures representation of both typical cloud environments and a wide variety of benign scripts that real cloud users might run.

\textbf{Malicious Set}: We collected malicious samples from VT and our HP. VT samples, with their first detection occurring no earlier than 2019 and their last detection in late 2024, were included if flagged as malicious by at least 10 out of 64 AV vendors. This threshold addresses limitations of traditional AVs in detecting script malware, which often prioritize low FPR over high TPR. We consider this multi-AV consensus a more reliable indicator of maliciousness, representing agreement among high-accuracy AVs such as Kaspersky, ESET-NOD32, FireEye, and etc.
Furthermore, we curated data from our HP, which enables to gather malicious samples from attackers directly targeting the cloud environment, and its samples have been verified as malicious via VT or by manually inspection if they are not in VT. SHA256 of example scripts from the HP are presented in Appendix \ref{apx:honeypot}. 
Our HP receives attacks from external sources, and a portion of those have VT reports while some do not appear on VT. Of the 3750 scripts in the malicious test set, 2911 have VT reports, while the remainder were manually verified as malicious.
A detailed breakdown of threat families and threat types of our malicious training and test sets—with VT reports—are presented in Appendix \ref{apx:threat_family} and \ref{apx:threat_type}.

\begin{figure}[pt]
\center
    \subfloat[]{\label{fig:length_plot}\includegraphics[width=0.48\textwidth]{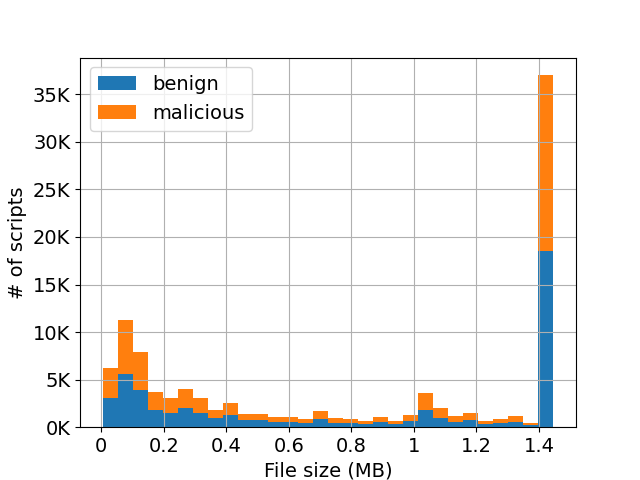}}
    \subfloat[]{\label{fig:entropy_plot}\includegraphics[width=0.48\textwidth]{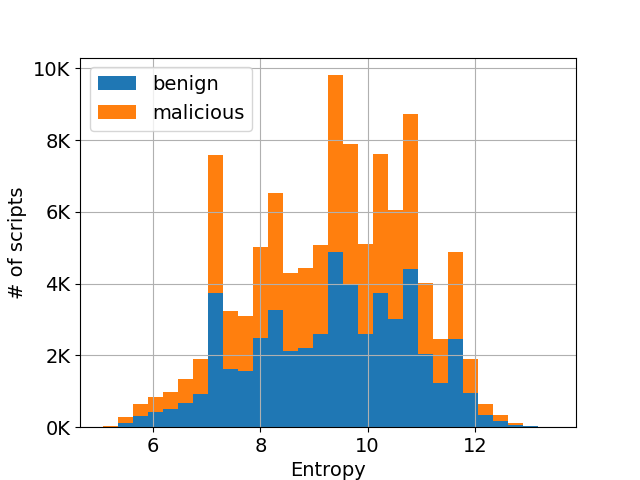}}
    \caption{Number of scripts with respect to (a) file size and (b) byte entropy in the training set.}
    \label{fig:dataset_stats}
\end{figure}

\textbf{Properties}: Our training dataset comprises a significant number of large files—specifically 40\% containing files larger than 1.2MB. The majority of these, constituting 60\%, are malicious and primarily sourced from VT and some from HP (see Figure \ref{fig:length_plot}). The presence of such large files is noteworthy as they may indicate obfuscation or a different format, as frequently observed in malicious cases. In Figure \ref{fig:entropy_plot}, we delve into the entropy of both benign and malicious scripts from the training set. Entropy serves as a metric for measuring randomness or disorder within data. Elevated entropy levels can hint at encryption, compression, or packing—common methods employed in obfuscation. Due to high volume of malicious files surpassing the mean entropy value of 9.5, Figure \ref{fig:entropy_plot} also shows that obfuscation is more enriched in our malicious scripts compared to benign. 

Large files and high entropy poses challenges for our approaches. Due to memory limitations, we introduce a cap on the number of tokens we can process from a file. Moreover, our parsing methods treat obfuscated code chunks as a single entity instead of de-obfuscating. However, it is imperative to evaluate the efficacy of our approaches in detecting malicious behavior, even when working with capped features and simple representation of obfuscations. This practice ensures a comprehensive assessment of our detection capabilities across diverse obfuscation scenarios.

\subsection{DL Model parameters}
To reduce the memory requirement for our models, we introduce a cap at $2048$ for the number of scope-token pairs for SCORE-H. Similarly, we limit the number of AST nodes of a script to $700$ and the length of their byte-strings to $512$ for SCORE-T. These limits are empirically selected for our dataset; larger sentence limits cause excessive zero padding of short scripts, which reduces the information content.
Our SM uses binary cross entropy loss, Adam optimizer with $lr = 0.001$ and weight decay$=0.0001$ for training, and GRL-based model uses hamming loss, Adam optimizer with $lr=0.0001$ and weight decay$=0.00001$ for learning the embeddings. We set the number of graph propagation to $T=5$.
The numerical results show performance comparisons between the proposed models, BitDefender, ClamAV and DL-based approaches: 1) MalConv2, which is a popular DL-based malware detector \cite{MalConv2}, trained with byte-level features (no preprocessing) as well as SCORE-T (BFT) features for a fair comparison, 2) SAST, which is a sequential model utilizing bi-LSTMs and AST-based features, 3) GAST, which is a GCN-based model utilizing graphical structure of AST features through their adjacency matrix, 4) UAST, which is a unified model comprising of SAST and GAST, and 5) an XGBoost detector that uses script embeddings from pre-trained CodeBERT model \cite{feng-2020-codebert}. We run all our experiments on an Intel(R) Xeon(R) CPU E5-2686 v4 @ 2.30GHz with 4 Tesla V100-SXM2 GPUS.

\begin{table*}[pt]
\centering
\caption{Performance results for ClamAV, BitDefender, MalConv2 \cite{MalConv2}, SAST, GAST and UAST \cite{UAST} and the proposed models in terms of F1 score, FPR, TPR/Recall, Precision and AUROC score.}
\label{tab:performance_comparison_all}
\setlength{\tabcolsep}{8pt}
\scalebox{0.85}{
\begin{tabular}{l|c|c|c|c|c|c|c}
\toprule
\textbf{Malware Detectors} & \textbf{Features}  & \textbf{Variations} & \textbf{F1} & \textbf{FPR}  & \textbf{Precision} & \textbf{TPR/Recall}  & \textbf{AUROC} \\ \midrule
ClamAV & Files & $-$ & $0.28993$ & $0$ & $1$ & $0.16954$ & $-$ \\ \midrule
        BitDefender & Files &  $-$  & $0.85590$ & $0$ & $1$ & $0.74810$ & $-$  \\ \midrule
        \cite{MalConv2} MalConv2 & Byte-strings & $-$ & $0.97016$ & $0.00188$ & $0.99813$ & $0.94371$ & $0.99756$  \\ \midrule
        \cite{MalConv2} MalConv2 & SCORE-T & BFT & $0.97350$ & $0.00180$ & $0.99810$ & $0.95010$ & $0.99784$  \\ \midrule
        \cite{UAST} SAST  & AST & BFT & $0.95276$ & $0.00184$ & $0.99791$ & $0.91152$ & $0.99781$ \\ \midrule
        \cite{UAST} GAST & AST & - & $0.69786$ & $0.00192$ & $0.99642$ & $0.53698$ & $0.99443$ \\ \midrule
        \cite{UAST} UAST & AST & BFT & $0.95967$ & $0.00185$ & $0.99813$ & $0.92391$ & $0.99690$ \\ \midrule
        CodeBERT XGBoost & CodeBERT embed. & $-$ & $0.96665$ & $0.00192$ & $0.99795$ & $0.93725$ & $0.99060$  \\ \midrule
        $[$Ours$]$ SM & SCORE-H & $-$ & $0.97732$ & $0.00185$ & $0.99805$ & $0.95744$ & $0.99922$  \\ \midrule
        \multirow{2}{*}{$[$Ours$]$ \textbf{SM}} & \multirow{2}{*}{\textbf{SCORE-T}} & DFT & $0.97932$  & $0.00180$ & $0.99806$ & $0.96128$ & $0.99909$  \\
        & &\textbf{BFT} & $\mathbf{0.98949}$  & $\mathbf{0.00172}$     & $\mathbf{0.99825}$      & $\mathbf{0.98088}$    & $\mathbf{0.99909}$ \\  \midrule  
        \multirow{2}{*}{$[$Ours$]$ GRL} & \multirow{2}{*}{SCORE-T} & Label-wise  & $0.98175$ & $0.00185$ & $0.99807$ & $0.96595$    & $0.99917$  \\
        & & Threat-wise & $0.98823$ & $0.00185$ & $0.99810$ & $0.97855$ & $0.99901$  \\ \bottomrule
\end{tabular}}
\end{table*}

\subsection{Evaluation}
Here, we present experimental results for our script malware detectors: SM with SCORE-H, SM with SCORE-T and GRL with SCORE-T, as well as their comparisons with AVs, i.e., BitDefender and ClamAV, and DL-based detectors, i.e., MalConv2 \cite{MalConv2}, SAST, GAST and UAST \cite{UAST}, and a CodeBERT-based detector. Table \ref{tab:performance_comparison_all} shows their performances on our dataset. Throughout evaluations, we refer to our approaches by their malware detectors and feature extractors: SM with SCORE-H, SM with SCORE-T (DFT), SM with SCORE-T (BFT), GRL with SCORE-T (label-wise) and GRL with SCORE-T (threat-wise). 

\subsubsection{Overall Detection Results}
\label{subsec:Overall}
Overall, we observe a high detection rate which is greater than $0.95$ TPR for all the proposed approaches for a fixed FPR around $0.0017$. Both SM and GRL models with SCORE-T features outperform the SM with SCORE-H. This performance gap can be attributed to the effectiveness of ASTs in capturing intricate hierarchical script structure, surpassing the sequential syntax highlighting employed by SCORE-H. Our performance analysis, as depicted in Table \ref{tab:performance_comparison_all}, reveals notable distinctions among malware detectors concerning AST traversal methods for SM and contrastive learning strategies for GRL. SM with SCORE-T has better performance when the AST is traversed in BFT order than in DFT order. Here, the advantage of BFT becomes evident due to the AST node cap, allowing for broader coverage of oversized scripts compared to DFT. This proves advantageous when malicious behavior manifests later in the code, enhancing the model's capability to detect threats effectively.
Furthermore, GRL with SCORE-T demonstrates enhanced performance when leveraging threat-wise clustering for contrastive learning of graph-based embeddings, as opposed to label-wise clustering. Contrastive learning, which bases representation learning on script similarity, benefits significantly from syntactic structural similarity provided by threat identification in contrast to the binary malicious-or-benign information provided by label-wise clustering. Thus, GRL with SCORE-T showcases superior performance in threat-wise contrastive learning, enhancing its ability to discern and detect threats effectively compared to the label-wise case.

\subsubsection{FP and FN Analysis}
\label{subsec:FPFN}
Upon conducting an analysis of FPs and FNs detected by our approaches, we identified common causes for FPs. The majority of FPs were attributed to two main factors: 1) manually written data in scripts and 2) the presence of non-English characters. Long strings of hard-coded numbers lead to confusion in the detector as it mimicks obfuscation. Additionally, non-English characters are often used to hide suspicious behavior in the code or pass language-specific security controls of signature based anti-virus tools. Since our training set potentially contains these types of malware, all our approaches including the AST-based models mis-classified these scripts as malicious. Following this intuition, we expect this information to be coming from the byte features since the AST nodes only contain the hierarchical syntactic structure of the code rather than the content of byte-strings. Therefore, this is relatively easy to mitigate for all approaches by introducing weights between the features from AST or syntax highlighting and the byte-features paired with them.

Majority of FNs were resulted from feature cap introduced due to memory constraints. This cap on feature length might limit capturing the entirety of script malware depending on the file size and AST traversal order, leading to instances where malicious behavior occurring later in a large script is missed. On the other hand this is partially mitigated by 1) choosing BFT over DFT order for traversing the ASTs, which allowed us to prioritize capturing the overall script structure in the parent nodes first before delving into the details of child nodes, and 2) increasing the feature cap size. While the former reduced FNs of Bash scripts significantly, the latter improved FNs of the rest of the languages as well. Another way to mitigate FNs can be introducing a sliding window of features over the entire file length and concatenating them before passing thorough CNN-based hierarchical embedding extractors of both SCORE-H and SCORE-T.

\subsection{Comparison with AVs}
\label{subsec:ComparisonAV}
We compare our approach with traditional AV solutions to evaluate its effectiveness against common defenses in cloud environments.
Our findings in Table \ref{tab:performance_comparison_all} reveal that our proposed approaches exhibit a notable improvement in detection rates when comparing our proposed approaches to traditional AV solutions ClamAV\footnote{\url{https://www.clamav.net/}} and BitDefender\footnote{\url{https://www.bitdefender.com/}}. These are widely recognized AV solutions in the cybersecurity domain. We selected these AVs for comparison due to their robust Linux support (since most clouds are based on Linux), their capability for script detection, and their widespread use in cloud environments. ClamAV is a popular open-source antivirus that runs natively on Linux and is known for detecting various malware types in scripts, while BitDefender, a commercial AV product, offers Linux support and advanced features with its comprehensive database of signatures allowing it to effectively detect and mitigate a wide range of threats. 

This improvement of our models is particularly noteworthy given a minor increase in FPR. The enhanced performance of our methods stems from their behavior-based detection mechanisms, which go beyond the limitations of static signature-based approaches. SCORE-T and SCORE-H are designed to extract the syntactic structure of malicious scripts, while our SM and GRL detectors are trained to learn the contextual nuances of these representations within the dataset. This performance gap is apparent particularly because of irregular and complex structure of scripts which are better captured by our behavior-based detection techniques. This enables our approaches to effectively identify and mitigate threats that may evade traditional signature-based AV solutions.

\subsection{Comparison with DL-based Approaches}
\label{subsec:ComparisonDL}
Our proposed models are evaluated against various DL-based approaches to assess their quality from multiple angles. We consider the following approaches for comparison, namely 1) MalConv2 \cite{MalConv2}, a malware detector that employs CNNs and trained with byte-level features, 2) MalConv2 trained with SCORE-T (BFT) features treated as byte-strings, 3) AST-based script classifiers of \cite{UAST} that utilize a sequential model with attention (SAST), a GCN-based model (GAST) and their joint model (UAST), and 4) an XGBoost-based detector that is trained with CodeBERT embeddings. 
Firstly, the ablation study comparing our approaches to (1) and (2) highlights the benefits of incorporating syntactic structures into script malware representations over byte-level representations. Additionally, comparisons with other AST and graph-based script representations (3) in the literature demonstrate the effectiveness of our approaches. These representations, i.e., SAST, GAST and UAST, are originally designed for task classification, and repurposed to detect malware in our comparisons. Finally, our comparison with CodeBERT representations demonstrate the superiority of syntax structure over more popular foundational model based feature representations. These decisions were driven by our goal to showcase how well our approaches understand code behavior within the context of script malware, and our aim is to demonstrate the superiority of our methods, even when compared to other code representation learning approaches in addition to malware detectors.

As shown in Table \ref{tab:performance_comparison_all}, most DL-based models demonstrate superior TPR compared to ClamAV and BitDefender, with a slight increase in FPR. However, GAST \cite{UAST} deviates from this trend, exhibiting weaker performance in the low FPR range despite comparable AUROC. This suggests GAST's reliance on GCNs, capturing only spatial correlations between AST nodes, limits its effectiveness in our target operating range compared to the spatio-temporal analysis of SM-based approaches using LSTMs. Notably, MalConv2 excels among DL-based models, surpassing SAST and UAST in TPR due to its ability to directly process large files as raw byte-strings. Performance further improves when MalConv2 is trained on SCORE-T (BFT) features treated as byte-strings, highlighting the efficacy of SCORE-T feature extraction. Furthermore, integrating GRL with SCORE-T significantly benefits both graph and sequence-based models from \cite{UAST} and MalConv2. Notably, threat-wise similarity learning in GRL achieves a remarkable TPR of $0.97855$, demonstrating the advantage of incorporating threat labels into similarity calculations. The superior performance of GRL over CodeBERT-XGBoost, using identical XGBoost parameters for both, highlights the richer information content provided by GRL's graphical representation of byte-strings and ASTs.

Moving to sequential models, SM with SCORE-T (BFT) emerges as the top performer with an impressive TPR of $0.98088$. Even though, the performance gap between SM with SCORE-T (BFT) and GRL with SCORE-T (therat-wise) is small, in situations where threat labels are absent or where minimizing FPR is paramount, SM with SCORE-T (BFT) demonstrates the most favorable trade-off between FPR and TPR. Several factors contribute to the success of this approach over the rest, including the efficacy of SCORE-T features in capturing script behavior compared to solely relying on AST features as seen in \cite{UAST}, our hierarchical embedding extraction method, and the effectiveness of BFT serialization. Additionally, the advantage of BFT becomes evident due to the AST node cap, allowing for broader coverage of oversized scripts compared to DFT. This proves advantageous when malicious behavior manifests later in the code, enhancing the model's capability to detect threats effectively.

While our models exhibit significant superiority over AST-based approaches in \cite{UAST}, the performance gap between MalConv2 and our models remains relatively narrow, primarily due to the constraints imposed by the AST node cap. MalConv2's ability to process a larger segment of the scripts confers a notable advantage both for MalConv2 byte-level and MalConv2 SCORE-T (BFT) since it treats both inputs as byte-strings.
In summary, our findings underscore the efficacy of SCORE-T features in conveying script behavior across various scenarios. Additionally, sequential modeling with hierarchical embeddings extracted from BFT features enhances the ability to obtain sophisticated structural information from scripts, further contributing to its performance advantage.

\begin{table}[pt]
\centering
\caption{Threat coverage of SM with SCORE-T (BFT) on test set (VT threat families). Singleton means AVClass \cite{AVClass} cannot identify a family name, VT Unknown are not found in VT but manually identified as malicious, and VT Missed are labeled as benign by VT but identified as malicious by our model and manual inspection.}
\vspace{0.2cm}
\label{tab:threat_family_VT}
\setlength{\tabcolsep}{8pt}
\scalebox{0.9}{
\begin{tabular}{l|c|c|c|l|c|c|c}
\toprule
\textbf{Family} & \textbf{Success} & \textbf{Miss} & \textbf{Ratio\%} & \textbf{Family} & \textbf{Success} & \textbf{Miss} & \textbf{Ratio\%} \\
\midrule
singleton     & 842 & 15 & 98.25 & surfbuyer   & 8  & 0  & 100 \\ \midrule
bundlore      & 791 & 2  & 99.75 & dstealer    & 6  & 0  & 100 \\ \midrule
bash          & 308 & 0  & 100   & loudminer   & 6  & 0  & 100 \\ \midrule
meterpreter   & 108 & 0  & 100   & shellbot    & 6  & 0  & 100 \\ \midrule
medusalocker  & 74  & 0  & 100   & cimpliads   & 5  & 1  & 83.33 \\ \midrule
disco         & 42  & 1  & 97.67 & rocke       & 4  & 0  & 100 \\ \midrule
mackeeper     & 39  & 1  & 97.5  & necrobot    & 4  & 0  & 100 \\ \midrule
pykeylogger   & 34  & 0  & 100   & tirrip      & 4  & 1  & 80 \\ \midrule
rozena        & 33  & 1  & 97.06 & pypws       & 4  & 0  & 100 \\ \midrule
shlayer       & 32  & 0  & 100   & chopper     & 4  & 0  & 100 \\ \midrule
mirai         & 29  & 0  & 100   & mccrash     & 4  & 0  & 100 \\  \midrule
memlod        & 19  & 0  & 100   & veil        & 3  & 0  & 100 \\ \midrule
dispread      & 16  & 0  & 100   & lazagne     & 3  & 0  & 100 \\ \midrule
pbot          & 14  & 0  & 100   & yellowdye   & 0 & 2   & 0 \\  \midrule
morila        & 10  & 0  & 100   & pyagent     & 0 & 2   & 0 \\ \midrule
kinsing       & 9   & 0  & 100   & remoteshell & 0 & 1   & 0 \\ \midrule
discorder     & 8   & 0  & 100   & c99shell    & 0 & 1   & 0 \\ \midrule
\textbf{VT Unknown}  & 1142 & 25 & 97.85 &  \textbf{VT Missed} & 79 & 7 & 91.86 \\
\bottomrule
\end{tabular}}
\end{table}

\subsection{Threat coverage and robustness}
\label{subsec:ThreatCoverage}
We delve deeper into the threat coverage of our top-performing model, i.e., SM with SCORE-T (BFT). Table \ref{tab:threat_family_VT} and \ref{tab:threat_coverage_VT} show threat family and threat type coverage of SM with SCORE-T (BFT), respectively, for top 30 threats in the test set. These threats are classified according to VT threat-naming, which incorporates 64 vendors in addition to BitDefender. Threat labeling is performed using AVClass \cite{AVClass}, which provides a label based on a consensus among multiple AV vendors. Two special categories are worth noting: VT Unknown and VT Missed.
\textbf{VT Unknown} represents the malicious samples that were not found in VT database but manually identified as malicious. 
These scripts are sourced from our HP and they have not been uploaded to/scanned by VT. Moreover, \textbf{VT Missed} represents the scripts that were labeled as benign by VT but identified as malicious by our model as well as manual inspection. An example from these scripts is as follows:
\begin{lrbox}{\mybox}%
\begin{lstlisting}[language=bash]
reset
cd /tmp && curl -O https://raw.githubusercontent.com/robertreynolds2/mine/main/mina.tar && tar -xf mina.tar && rm -rf mina.tar && chmod +x update logs && bash ./logs > /dev/null 2>&1
\end{lstlisting}
\end{lrbox}
\begin{center}
\hspace*{0.75cm}\scalebox{0.9}{\usebox{\mybox}}
\end{center}
This script demonstrates a stealthy attack method. It downloads a payload from a untrusted repository to the /tmp directory, extracts it, and immediately removes the original archive. The script then executes the extracted content, suppressing all output to avoid detection. Our SM with SCORE-T (BFT) model detected this script as malicious, which VT missed.  As a result of our manual inspection, we determine that this script is malicious.
\begin{lrbox}{\yrbox}%
\begin{lstlisting}[language=bash]
wget http://136.144.41.46/b/b.x86
chmod 777 b.x86
./b.mips Apache
rm -rf b.x86
rm -rf $0
\end{lstlisting}
\end{lrbox}
\begin{center}
\hspace{-9cm}\scalebox{0.9}{\usebox{\yrbox}}
\end{center}
Similarly, this script exemplifies a typical botnet malware downloader, likely associated with Mirai or Gafgyt families. It downloads an x86 binary from an untrusted IP address, grants it full permissions, and then executes a MIPS binary with "Apache" as an argument. This execution method suggests potential process injection or an attempt to disguise the malware as a legitimate Apache process. The script then removes both the downloaded file and itself to evade detection. Our SM with SCORE-T (BFT) model successfully identified this script as malicious, which VT missed. Our manual inspection confirmed its maliciousness, highlighting the script's use of multi-architecture payloads and evasion techniques common in botnet operations.

\begin{table}[pt]
\centering
\caption{Threat coverage of SM with SCORE-T (BFT) on the Test Set (VT threat types). VT Unknown are not found in VT but manually identified as malicious.}
\vspace{0.2cm}
\setlength{\tabcolsep}{8pt}
\scalebox{0.9}{
\begin{tabular}{l|c|c|c|l|c|c|c}
\toprule
\textbf{Threat Type} & \textbf{Success} & \textbf{Miss} & \textbf{Ratio\%} & \textbf{Threat Type} & \textbf{Success} & \textbf{Miss} & \textbf{Ratio\%} \\
\midrule
downloader    & 940 & 4  & 99.58 & keylogger     & 11  & 0  & 100  \\ \midrule
grayware      & 907 & 6  & 99.34 & grayware:tool & 11  & 0  & 100  \\ \midrule
backdoor      & 146 & 5  & 96.69 & expkit        & 10  & 0  & 100  \\ \midrule
worm          & 101 & 1  & 99.02 & bot           & 6   & 0  & 100  \\ \midrule
cryptominer   & 95  & 3  & 96.94 & webshell      & 6   & 1  & 85.71 \\ \midrule
\textbf{VT Unknown} & 1895 & 42 & 97.83 & ransomware & 6   & 1  & 85.71 \\
\bottomrule
\end{tabular}}
\label{tab:threat_coverage_VT}
\end{table}
Table \ref{tab:threat_coverage_VT} shows that our model achieves full or at least 90\% coverage for most threat families including the scripts missed by VT. Although SM with SCORE-T (BFT) exhibits lower coverage for yellowdye, pyagent, remoteshell and c99shell, overall it shows a high threat coverage for significant threat types and detects a reasonable number of threats that VT vendors missed. Table \ref{tab:threat_coverage_VT} also shows that our model's coverage for critical threat types, such as backdoor, cryptominer, webshell and ransomware, is significantly high.

To evaluate the robustness of our model against obfuscated scripts, we examined common obfuscation techniques in our test set, including XOR encryption, ROT13 substitution cipher, Base64 encoding, and PowerShell Encoded Commands. Table \ref{tab:robustness_summary} details the distribution of these potentially obfuscated samples, with ROT13 encryption and Base64 encoding being particularly prevalent. SM with SCORE-T (BFS) accurately detects the majority of obfuscated samples with the correct label. Despite not decoding the obfuscated parts during AST construction, our model still identifies XOR, ROT13, and Base64 obfuscations highly accurately, as well as PowerShell Encoded Commands effectively. This demonstrates the resilience of our model against common obfuscation techniques.

\begin{table}[pt]
\centering
\caption{Count of obfuscated samples in test set and it's ratio to total number of scripts. SM with SCORE-T (BFS) predicted obfuscated samples with high accuracy.}
\vspace{0.2cm}
\setlength{\tabcolsep}{8pt}
\resizebox{0.6\linewidth}{!}{%
\begin{tabular}{l|c|c|c|l|c|c|c}
\toprule
\textbf{Obfuscation} & \textbf{Count} & \textbf{Ratio\%} & \textbf{Accuracy} \\ \midrule
Base64 Encoding  & 3349 & 44.65 & 98.38 \\ \midrule
XOR Encryption   & 5    & 0.067 & 100   \\ \midrule
ROT13 Encryption & 6937 & 92.49 & 98.94 \\ \midrule
PowerShell Encoded Command  & 17 & 0.227 & 70.58 \\ \bottomrule 
\end{tabular}}
\label{tab:robustness_summary}
\end{table}

\section{Discussion}
\noindent\textbf{Extensibility to Other Script Languages.} Although our approach is designed specifically for server-side scripting languages such as Python, Perl, and Bash, the design principles do not rely on specific features of these languages. SCORE-H, which represents syntactic functionality as sequences, and SCORE-T, which captures deeper understanding through hierarchical syntactic representation, can be adapted to other scripting languages with syntax and structure. Although this may seem ambitious, it's essential to note that our approach leverages the inherent syntax and structure of scripts, making it possible to apply the same preprocessing and DL-based detection to other languages with similar characteristics. Our method requires only a parser as a language specific tokenizer. The integration of these extractors with a sequential neural network and the incorporation of hierarchical structure into embeddings through GRL are generalizable techniques. Adapting our approach to a new language mainly involves developing language-specific extractors, collecting representative samples, and training the models accordingly. This extensibility demonstrates the potential for our approach to have a wider impact on detecting script malware threats beyond the server-side languages explored in this paper, such as client-side languages like JavaScript.

\noindent\textbf{Resilience Against Obfuscation.} 
While our approaches demonstrate promising results in detecting server-side scripting malware, there are some limitations to be considered. Both SCORE-H and SCORE-T treat obfuscated code as a single code-chunk or token without processing their corresponding byte-strings. During the training phase, our models learn whether the extracted features from the obfuscated scripts in the training set are malicious or benign, rather than de-obfuscating the code. However, leveraging graph representations, structure based feature extraction methods like ASTs and syntax highlighting provide a certain level of resilience against common obfuscation techniques such as string encryption/encoding, variable and function renaming, dead code insertion. This is presented and discussed in Section \ref{subsec:ThreatCoverage} for Base64, XOR, ROT13, and PowerShell encodings. Although our approaches do not decode them, they still identify Base64, XOR, ROT13 obfuscations accurately using the hierarchical syntactic structure and bytestrings of the scripts and learning their spatial characteristics.

While no system is perfect, our approaches face limitations when confronted with advanced obfuscation techniques. If a script is entirely packed or its logic heavily obfuscated through transformations disrupting data flow and control flow combined with string obfuscation, our code parsing-based extractors may struggle to generate meaningful features. In such cases where the obfuscation is extensive, our models could potentially treat the entire script as a single obfuscated token or misinterpret the code chunks, leading to misclassification. However, for server-side scripts like Python, Perl, and shell, the inherent syntactic structure of functions and commands themselves provide substantial discriminative information. This makes it challenging for these obfuscation techniques alone to evade our detection methods effectively. Therefore, while our current approaches have limitations when dealing with scripts that are completely packed or undergo heavy logical obfuscation, they demonstrate resilience against common obfuscation methods by leveraging the rich information present in interpreted code.

\noindent\textbf{Training Data Representativeness Challenges.} Training data quality is a crucial factor influencing the performance of our malware detection models. Our approaches heavily rely on the comprehensiveness and diversity of the benign and malicious scripts used for training, including coverage across multiple scripting languages. If certain functionality or script behaviors are underrepresented in the training set, it can lead to biased results and blind spots in detection. For instance, if the benign set lacks scripts exhibiting network communication behaviors, the model may incorrectly flag all such behaviors as malicious. Conversely, if the training data includes an overabundance of benign scripts resembling malicious functionality, such as credential harvesting, the model may fail to accurately identify these as threats. Achieving the right balance and representative coverage in the training data is essential to tailor our system's sensitivity towards the specific malicious behaviors and functionalities of interest. 

\noindent\textbf{Fixed Input Length.} The SCORE-H and SCORE-T models necessitate fixed-length inputs for their feature embedding modules, constraining the extent of script analysis. Our dataset comprises files ranging from 4.2 KB to 1.45 MB in length. For SCORE-T, handling this variability involves truncating AST nodes to a maximum limit, discarding surplus nodes beyond this threshold. The truncation eliminates different parts of scripts based on the traversal order, whether breadth-first or depth-first. Since AST node features are concatenated with their corresponding byte-strings, any truncated nodes also result in the elimination of associated bytestring information. Additionally, lengthy byte-strings combined with AST nodes undergo chunking, with excess byte-strings discarded.

Similarly, SCORE-H features, both syntactic and byte-level, are truncated when exceeding the specified feature cap. This fixed-length input constraint inevitably leads to information loss, especially for longer scripts. While most malicious files tend to be relatively short, there is a potential risk of evasion for longer files containing small malicious segments amidst predominantly benign code due to truncation. To mitigate this limitation, we are exploring a sliding window-based feature extraction approach as part of future work, enhancing our model's capability to detect malicious segments within longer scripts.

\section{Conclusions}
In this paper, we introduce novel approaches that represent code as features and leverage DL models to extract meaningful information for the detection of malicious behavior. Our methods encompass parsing code using two distinct techniques: syntax highlighting (SCORE-H) and abstract syntax trees (SCORE-T). Additionally, we propose an SM and a GRL-based model designed to identify malicious behavior based on these features.
Our experimental results demonstrate the superiority of SM with SCORE-H over benchmark models such as MalConv2, CodeBERT representations, as well as the AST-based models in the literature, e.g., SAST, GAST, and UAST utilizing sequential and graph-based networks. Furthermore, we observe that SCORE-T features, with their enriched hierarchical structure information, lead to higher detection rates for both SM and GRL-based detectors, with the former excelling particularly when AST feature serialization is performed in BFT order due to long scripts and the latter excelling when we have access to threat family labels.
The comparative analysis reveals that, in addition to byte-string and token-based representations, our approaches significantly outperform AST-based models (SAST, GAST, and UAST) despite some similarities in their feature extractions. This underscores the effectiveness of our joint byte-syntactic features, leveraging hierarchical embedding extraction to provide superior script representations for malware detection.
Overall, our experiments demonstrate that the SM processing hierarchical embeddings from AST representations of scripts in BFT order, as in SCORE-T, effectively capture and predict malicious behavior, while also offering a more favorable TPR-FPR trade-off compared to other approaches.

\newpage
\bibliographystyle{unsrt}
\bibliography{references} 

\newpage
\appendix

\section{Dataset Analysis}
In this section, we provide further details about our training and test datasets. We obtain threat family and threat type distributions from VT reports using AVClasss package \cite{AVClass}, which we utilize for threat labeling. Moreover, we present exemplar scripts from our honeypot that do not appear in VT database.
\subsection{Threat Family}
\label{apx:threat_family}
We investigated the threat family distribution for our training and test datasets that appears in VT database. Table \ref{tab:threat_family_train_data} and \ref{tab:threat_family_test_data} show top 30 threat families from training and test set, respectively. \textit{Singleton} on both tables mean AVClass package \cite{AVClass} cannot identify a family name for a script. Both tables exhibit a wide variety of threat families offering a clearer understanding of the distribution within our in-house dataset.

\begin{table}[pb]
\centering
\caption{Top 30 threat families, their counts and ratios (\%) in the training dataset. }
\vspace{0.3cm}
\setlength{\tabcolsep}{8pt}
\scalebox{0.9}{
\begin{tabular}{l|r|r|l|r|r}
\toprule
\textbf{Family} & \textbf{Count} & \textbf{Ratio\%} & \textbf{Family} & \textbf{Count} & \textbf{Ratio\%} \\
\midrule
singleton     &10027 & 34.35  & morila    & 108 & 0.37 \\ \midrule
bundlore      & 8669 & 29.70  & surfbuyer & 107 & 0.37 \\ \midrule
bash          & 3345 & 11.46  & lazagne   & 100 & 0.34 \\ \midrule
meterpreter   & 1097 &  3.76  & discorder & 96  & 0.33 \\ \midrule
medusalocker  & 807  &  2.76  & chopper   & 71  & 0.24 \\ \midrule
pykeylogger   & 565  &  1.93  & kinsing   & 66  & 0.23 \\ \midrule
disco         & 487  &  1.67  & veil      & 60  & 0.21 \\ \midrule
mackeeper     & 470  &  1.61  & loudminer & 53  & 0.18 \\ \midrule
mirai         & 414  &  1.42  & tirrip    & 51  & 0.17 \\ \midrule
rozena        & 345  &  1.18  & dstealer  & 46  & 0.16 \\ \midrule
shlayer       & 253  &  0.87  & mccrash   & 43  & 0.15 \\ \midrule
shellbot      & 221  &  0.76  & pypws     & 43  & 0.15 \\ \midrule
memlod        & 206  &  0.71  & dakkatoni & 30  & 0.10 \\ \midrule
dispread      & 199  &  0.68  & rocke     & 30  & 0.10 \\ \midrule
pbot          & 139  &  0.48  & minerdownloader & 30 & 0.10 \\
\bottomrule
\end{tabular}}
\label{tab:threat_family_train_data}
\end{table}

\begin{table}[pt]
\centering
\caption{Top 30 threat families, their counts and ratios (\%) in the test dataset. }
\vspace{0.3cm}
\setlength{\tabcolsep}{8pt}
\scalebox{0.9}{
\begin{tabular}{l|r|r|l|r|r}
\toprule
\textbf{Family} & \textbf{Count} & \textbf{Ratio\%} & \textbf{Family} & \textbf{Count} & \textbf{Ratio\%} \\
\midrule
singleton     &1010 & 34.70  & surfbuyer     & 11 & 0.38 \\ \midrule
bundlore      & 865 & 29.71  & morila        &  8 & 0.27 \\ \midrule
bash          & 322 & 11.06  & kinsing       &  8 & 0.27 \\ \midrule
meterpreter   & 123 &  4.22  & lazagne       & 8  & 0.27 \\ \midrule
medusalocker  & 88  &  3.02  & tirrip        & 7  & 0.24 \\ \midrule
disco         & 54  &  1.86  & discorder     & 7  & 0.24 \\ \midrule
pykeylogger   & 45  &  1.55  & pykeylog      & 6  & 0.21 \\ \midrule
mirai         & 39  &  1.34  & chopper       & 6  & 0.21 \\ \midrule
mackeeper     & 33  &  1.13  & jackshell     & 5  & 0.17 \\ \midrule
shlayer       & 31  &  1.06  & pypws         & 5  & 0.17 \\ \midrule
rozena        & 27  &  0.93  & remoteshell   & 5  & 0.17 \\ \midrule
pbot          & 22  &  0.76  & veil          & 5  & 0.17 \\ \midrule
shellbot      & 22  &  0.76  & dstealer      & 5  & 0.17 \\ \midrule
memlod        & 18  &  0.62  & mccrash       & 5  & 0.17 \\ \midrule
dispread      & 17  &  0.58  & minerdownloader & 5 & 0.17 \\ 
\bottomrule
\end{tabular}}
\label{tab:threat_family_test_data}
\end{table}

\subsection{Threat Type}
\label{apx:threat_type}
We present the various threat types found in our training and test sets in Table \ref{tab:threat_types_train_data} and Table \ref{tab:threat_types_test_data}, respectively. These tables illustrate a diverse range of threat types, with a significant portion being downloaders and a substantial number consisting of critical threats such as backdoors, coinminers, ransomware, and webshells.

\begin{table}[pt]
\centering
\caption{Threat types, their counts and ratios (\%) in the training dataset. No threat types were found for 3.7k samples.}
\vspace{0.3cm}
\setlength{\tabcolsep}{8pt}
\scalebox{0.9}{
\begin{tabular}{l|r|r|l|r|r}
\toprule
\textbf{Threat Types} & \textbf{Count} & \textbf{Ratio\%} & \textbf{Threat Types} & \textbf{Count} & \textbf{Ratio\%} \\
\midrule
downloader  &10762 & 36.87\%  & tool        & 117 & 0.40 \\ \midrule
grayware    & 9931 & 34.03\%  & expkit      & 73  & 0.25 \\ \midrule
backdoor    & 1737 &  5.95\%  & ransomware  & 73  & 0.25 \\ \midrule
worm        & 1285 &  4.40\%  & webshell    & 56  & 0.17 \\ \midrule
miner       & 1161 &  3.98\%  & virus       & 11  & 0.04 \\ \midrule
bot         & 152  &  0.52\%  & rootkit     &  7  & 0.02 \\ \midrule
keylogger   & 125  &  0.43\%  & spyware     &  1  & 0.003 \\
\bottomrule
\end{tabular}}
\label{tab:threat_types_train_data}
\end{table}

\begin{table}[pt]
\centering
\caption{Threat types, their counts and ratios (\%) in the test dataset. No threat types were found for 3.7k samples.}
\vspace{0.3cm}
\setlength{\tabcolsep}{8pt}
\scalebox{0.9}{
\begin{tabular}{l|r|r|l|r|r}
\toprule
\textbf{Threat Types} & \textbf{Count} & \textbf{Ratio\%} & \textbf{Threat Types} & \textbf{Count} & \textbf{Ratio\%} \\
\midrule
downloader  & 1085 & 37.26  & bot         & 13  & 0.45 \\ \midrule
grayware    &  978 & 33.58  & tool        & 11  & 0.38 \\ \midrule
backdoor    &  205 &  7.04  & expkit      & 11  & 0.38 \\ \midrule
worm        &  134 &  4.60  & ransomware  & 11  & 0.38 \\ \midrule
miner       &  103 &  3.54  & webshell    & 5   & 0.17 \\
keylogger   &  18  &  0.62  & & & \\
\bottomrule
\end{tabular}}
\label{tab:threat_types_test_data}
\end{table}

\newcommand{\hash}[1]{{\ttfamily\seqsplit{#1}}}

\subsection{HoneyPot Script Examples}
\label{apx:honeypot}
In this section, we provide SHA256 and descriptions of example scripts from our honeypot to illustrate the malicious behavior of scripts, and demonstrate how we manually inspect and verify their malicious nature.

The Perl script with SHA256: \hash{2eaa3d77958e0f9b5c2725b504804ed2573097b941323ac44968abf01e8b00d0} is designed to flood a target IP address with UDP packets, potentially overwhelming the target's resources and making it unavailable to legitimate users. This is a form of Denial of Service (DoS) attack. The purpose of the script is to disrupt the operation of a network or a specific target, which is our honeypot in this case.









This Bash script with SHA256: \hash{0a5610b3d8ad9404820f50402e2903cc57417a410dc436f2840adac9805278ee} sets up and runs cryptominer software on a compromised system. It sets up persistent access through a cron job, ensuring it continues to run even if detected and stopped. It downloads and executes files from an untrusted source, and uses system resources to mine cryptocurrency. It attempts to hide its activities by running in the background and redirecting output.


The next Perl script with SHA256: \hash{6cf6e51cdf0fbde32feed7341601e0fb6eaf7d47cb32c6b9678899eb16628f5f} creates a reverse shell, allowing an attacker to remotely access and control the system it's run on. It works by establishing a TCP connection to a specified remote IP address and port, then redirecting the system's standard input, output, and error streams to this connection. This gives the attacker remote shell access to the compromised system, bypassing normal authentication and security measures. Once executed, it allows the attacker to run commands, access files, and potentially further exploit the system as if they had direct physical access. 


  
  
  
  
  
  
  
  
  

The next script with SHA256: \hash{0a218d7213d218d2756f5a35457a1f1a35cf1125fb3dd3fe811a079f5c37639c} downloads executables from a remote server and executes them without any verification. The script sets the permissions of the downloaded executables to 777, which means anyone can read, write, and execute them. The script attempts to download and execute binaries for various CPU architectures to infect a wide range of devices. The script also attempts to use busybox to download and execute the same files, which is often found in embedded Linux systems, indicating an attempt to compromise such devices. The last line (rm \$0) attempts to delete the script itself, which is a common tactic used by malware to cover its tracks.



\section{Script language coverage:}
\label{apx:language_coverage}
We assess the accuracy of SM with SCORE-T (BFT) across each programming languages included in our training set, i.e., Bash, Python and Perl. Table \ref{tab:lang_coverage} provides insights into the accuracy of SM with SCORE-T (BFT) concerning these languages across various datasets: the balanced test set (malicious-benign), the {\it Rest of Benign} set (comprising benign data not utilized in training), and the {\it VT benign} set (comprising files from VT with no malicious verdicts). In the test set, we observe lower accuracy for Perl. This is primarily due to the limited number of Perl scripts during training. However, this performance difference is less obvious in benign sets. It's essential to note that lack of malicious verdicts from VT does not guarantee that the scripts are benign. It's plausible that the dataset may contain unidentified malicious files that have not been flagged by any VT vendors. Hence, a slightly lower performance in the {\it VT benign} set compared to {\it Rest of Benign} is expected due to this uncertainty in the dataset's composition.

\begin{table}[pt]
\centering
\caption{Script language accuracy of SM with SCORE-T (BFT) on test set, rest of benign and VT benign sets.}
\vspace{0.3cm}
\label{tab:lang_coverage}
\scalebox{0.9}{
\setlength{\tabcolsep}{8pt}
\begin{tabular}{lccc}
\toprule
\textbf{Script Language}   & \textbf{Test Set} & \textbf{Rest of Benign} & \textbf{VT Benign}   \\ \midrule
 Bash       & 0.99119 & 0.99933 & 0.94921 \\ \midrule
 Python     & 0.97411 & 0.99952 & 0.91094 \\ \midrule
 Perl       & 0.96667 & 0.99107 & 1 \\
        \bottomrule
\end{tabular}}
\end{table}

\begin{table}[pt]
\centering
\caption{Average latency for ClamAV, BitDefender, MalConv2 \cite{MalConv2}, SAST, GAST and UAST \cite{UAST} and the proposed models per file in milliseconds.}
\vspace{0.3cm}
\label{tab:runtime}
\setlength{\tabcolsep}{8pt}
\scalebox{0.9}{
\begin{tabular}{lcc}
\toprule
\textbf{Malware detectors} & \textbf{Features}  & \textbf{Avg. latency / file} \\ \midrule
        ClamAV & Files & $<$ 1 ms  \\ 
        BitDefender & Files & $<$ 1 ms  \\ 
        \cite{MalConv2} MalConv2 & Bytes &  20 ms  \\ 
        \cite{UAST} SAST  & AST & 35 ms \\ 
        \cite{UAST} GAST & AST & 4 ms \\ 
        \cite{UAST} UAST & AST & 40 ms \\ 
        $[$Ours$]$ SM & SCORE-H & 50 ms \\ 
        $[$Ours$]$ SM & SCORE-T & 45 ms \\ 
        $[$Ours$]$ GRL & SCORE-T&  250 ms \\ 
        CodeBERT-XGBoost & CodeBERT embed. & 800 ms \\ \bottomrule
\end{tabular}}
\end{table}

\section{Latency analysis}
Table \ref{tab:runtime} presents a comparative analysis of latency performance among baseline models and proposed models, measured in average latency per file in milliseconds. These experiment are performed on an Intel(R) Xeon(R) CPU E5-2686 v4 @ 2.30GHz machine, utilizing only the CPU devices in PyTorch 2. As anticipated, BitDefender and ClamAV, functioning as rule-based AVs, emerge as the fastest malware detectors, closely followed by GAST---a simple graph convolutional network (GCN)---and MalConv2, which incorporates convolutional neural networks. Conversely, models utilizing recurrent neural networks (RNNs) exhibit latency ranging from 35 to 50 milliseconds, a result aligned with expectations given the increased complexity and slower processing inherent in RNN architectures compared to simpler convolutional methods. The GRL model, featuring graph similarity learning, introduces the first leap in the latency at 250 milliseconds per file due to its complex graph propagation algorithm. Finally, CodeBERT-XGBoost represents the pinnacle of complexity and consequently registers the largest latency at 800 milliseconds per file. Despite these disparities in latency efficiency, the DL-based models notably excel in accuracy when compared to BitDefender and ClamAV. This trade-off between latency and detection accuracy is a common consideration in the field of cybersecurity, where advanced DL models often sacrifice latency efficiency for improved detection capabilities. For tasks prioritizing swift processing of large volumes of files, a standard AV's minimal latency makes it an attractive choice, especially when real-time scanning and rapid response to threats are crucial. On the other hand, tasks that prioritize precise and comprehensive threat detection would find SM with SCORE-T slightly slower but much more accurate to be a better fit for cybersecurity needs.

\end{document}